# Digitization of Weather Records of Seungjeongwon Ilgi: A Historical Weather Dynamics Dataset of the Korean Peninsula in 1623–1910

Zeyu Lyu [1*], Kohei Ichikawa [2,3*], Yongchao Cheng [4], Hisashi Hayakawa [5], and Yukiko Kawamoto [6]

[1] Graduate School/Faculty of Arts and Letters, Tohoku University

[2] Frontier Research Institute for Interdisciplinary Sciences, Tohoku University

[3] Global Center for Science and Engineering, Faculty of Science and Engineering, Waseda University

[4] Center for Northeast Asian Studies, Tohoku University

[5] Institute for Advanced Research / Institute for Space–Earth Environmental Research, Nagoya University

[6] Graduate School of Humanities Department of Humanities, Nagoya University

*Corresponding Authors:

Zeyu Lyu: lyu.zeyu.e8@tohoku.ac.jp
Kohei Ichikawa: kohei.ichikawa@aoni.waseda.jp

**Abstract**

Historical weather records from Europe indicate that the Earth experienced substantial climate variability, which caused, for instance, the Little Ice Age and the global crisis in the period between the 14th and 19th centuries. However, it is still unclear how global this climate variability was because of the scarce meteorological data availability in other regions including East Asia, especially around the 17th century. In this context, *Seungjeongwon Ilgi*, a daily record of the Royal Secretariat of the Joseon Dynasty of Korea, is a precious source of historical meteorological records for the Korean Peninsula, as it covers 288 years of weather observations made during 1623–1910. We used the digital database of *Seungjeongwon Ilgi* to construct a machine-readable weather condition dataset. To this end, we extracted valid weather information from the original weather description text and compiled them into predefined weather categories. Additionally, we attempted to improve the usability of dataset by converting the reported dates in the traditional calendar system to those in the Gregorian calendar. Finally, we outlined promising implications of this dataset for meteorological and climatological studies, while describing the limitations of the dataset. Overall, future studies focusing on the climate and weather of the past could use this meteorological database for investigating long-term climate variability. Our datasets are publicly available at 10.5281/zenodo.8142701.

**Introduction**

It is widely acknowledged that the climate naturally changes at decadal to centennial timescales. Historical weather conditions have been abundantly recorded by numerous meteorological measurements over centuries, which has laid the foundation for understanding the long-term climate variabilities since the $17^{th}$–$18^{th}$ centuries (Brönnimann et al., 2019; Harrison, 2014; Jones et al., 2011). Several long-term and systematic efforts of meteorological observations have covered significantly long timescales with high time cadences and are often used for reconstructing the long-term climate variability. Such examples include the long-term set of meteorological measurements in England from 1658 onward, France from 1688 onward, and Ireland from 1716 onward (Brönnimann et al., 2019; Manley, 1974; Murphy et al., 2017; Parker et al., 1992; Slonosky, 2002). From a modern research perspective, such historical meteorological records have been subjected to digitization efforts as they provide unique perspectives on the long-term meteorological variability (Brönnimann et al., 2019; Hawkins et al., 2022).

From a climatological perspective, these records have valuable implications for local and global climate variability not only in the quasi-modern period but also for the Little Ice Age (LIA), which lasted from the $14^{th}$ to $19^{th}$ centuries (Grove, 1988; Matthes, 1939) and especially for the period of the global crises of the $17^{th}$ century (Parker, 2013). This period chronologically overlapped with the grand solar minimum and active volcanic eruptions (Hayakawa, Kuroyanagi, et al., 2021; Sigl et al., 2015), whereas the relationships with these variabilities were not somewhat controversial (Owens et al., 2017; Soon et al., 2014). However, it is challenging to quantify the climate variability in the $17^{th}$ century, because only few meteorological measurements offer a temporal coverage of this period. In this regards, the quantification of the past climate variability foremost requires digitization and the interpretation of early meteorological measurements (Domínguez-Castro et al., 2008; Pliemon et al., 2022; Slonosky, 2002).

Moreover, the aforementioned period has been studied using historical and proxy records ((Marcott et al., 2013; Oliva et al., 2018). In this context, East Asia is rich with historical diaries and chronicles in multiple fields, which also cover astronomical and meteorological observations (Aono and Saito, 2010; Hayakawa et al., 2017; Ichino and Masuda, 2022; Mizukoshi, 1993; Yang et al., 2019). One of such important long-term observations were conducted in Korea under Joseon Dynasty (1392–1910). *Seungjeongwon Ilgi* is a historical record of the Korean Peninsula, recording events within the palace and the country. This daily record, also stands out with unprecedented temporal coverage of weather observations, the period from 1623 to 1910 (288 years). This historical meteorological record laid the foundation for further meteorological and climatological studies, which relied on weather information, such as temperature and wind data. In particular, the relevant precipitation and temperature data have been previously analyzed and discussed by some researchers (Kim, 2016, 2018; Wada, 1917), but the weather data have remained unexplored.

In this study, we digitized weather records from the historical weather record *Seungjeongwon Ilgi* over 1623–1910 to transform these records into a machine-readable dataset. Such a dataset can be useful for scientists interested in long-term meteorological studies.

**Data Records**

*Seungjeongwon Ilgi* is a daily record of Seungjeongwon, the Royal Secretariat of the Joseon Dynasty of Korea (1392–1910). Seungjeongwon was an office of the king's secretary in charge of the nation's top secrets of this period; *Seungjeongwon Ilgi* recorded the public life of the king and his interactions with the bureaucracy on a daily basis, including the king's orders, documents, and incidents, in addition to reporting to the king and conveying royal decrees to the bureaucracy. *Seungjeongwon Ilgi* offers daily entries starting from daily representative weather reports, which notably include weather observations in the middle of entries in some cases. Figure 1 is a photo of the original copy of *Seungjeongwon Ilgi*, which is kept in Seoul National University Kyujanggak Institute for Korean Studies now. In this photo, we could find the weather report ("晴") after the date.

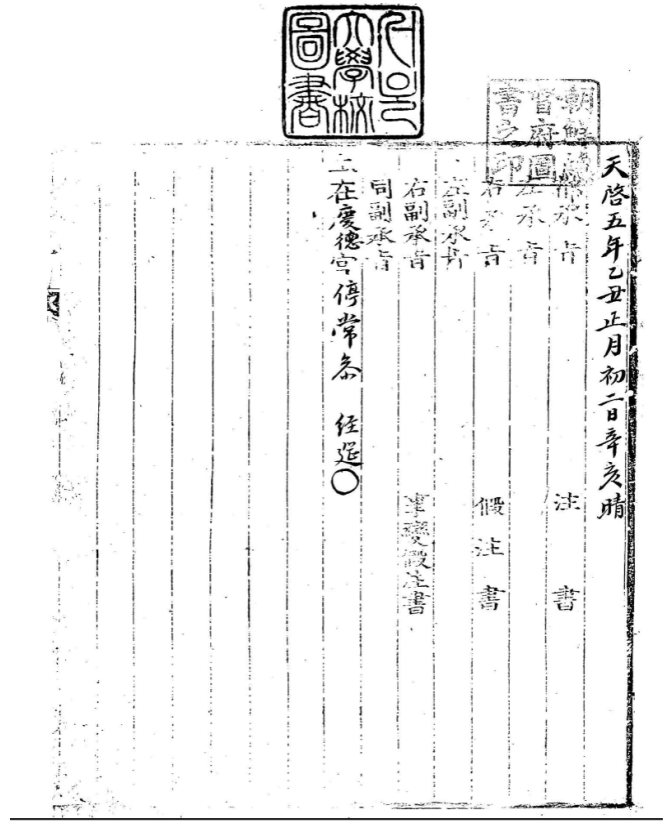

**Figure 1.** Photo of original *Seungjeongwon Ilgi* now preserved in Seoul National University Kyujanggak Institute for Korean Studies [1].

While the former type of records were interpreted by a historian, the latter records were reported by the scribes of *Seungjeongwon Ilgi* or meteorological observers (*Gwansanggam*), and a local governor (*Gwanchalsa*) (Kim, 2016). These records were accumulated from 1623 to 1910 in the form of daily weather reports at *Seungjeongwon* in the Korean palace complex (N 37°35′, E 126°59′) in modern Seoul (Donggwoldo, 1989; Hong, 2016). These records represent a priceless archive of data, reflecting actual meteorological conditions in Korea for 288 years.

Given the reform of the government and administrative systems, the name of the *Seungjeongwon* was changed several times from *Seungjeongwon* to *Seungseonwon*, *Gungnaebu*, *Biseogam*, and *Biseowon* after 1894. Thus, despite the titles of diaries have been called *Seungseonwon Ilgi*, *Gungnaebu Ilgi*, *Biseogam Ilgi*, or *Gyujanggak Ilgi* depending on the exact period, we hereafter call the entire royal diary collection *Seungjeongwon Ilgi* as this is a widely acknowledged annotation of these documents nowadays.

As *Seungjeongwon* was the largest top-secret state institution during the *Joseon* Dynasty, *Seungjeongwon Ilgi* is a primary source instead of the *Joseon Wangjo Sillok* (*the Annuals of the*

---

[1] https://kyudb.snu.ac.kr/

*Joseon Dynasty*) (NIKH, 1977). The immense value of *Seungjeongwon Ilgi* was recognized and registered as the Memory of the World in September 2001 by UNESCO (The United Nations Educational, Scientific and Cultural Organization). While the original manuscripts of *Seungjeongwon Ilgi* were written in a cursive script, which severely hampered their deciphering, their facsimile photocopies have been published by the National Institute of Korean History (NIKH) (NIKH, 1977). These records were consequently transcribed into text data and associated with their digital images in the database of NIKH[2]. Note that this database has also been published as the source dataset of *Seungjeongwon Ilgi*[3], which we used and analyzed in this study.

**Data Production Methods and Major Roadblocks**

This study scrutinized the NIKH database for *Seungjeongwon Ilgi* and digitized the daily weather records as a resource for assessing the meteorological and climatological variabilities in the Korean Peninsula at a centennial timescale. Although *Seungjeongwon Ilgi* provides useful daily weather records, it is difficult to utilize the digital weather dataset due to several challenges in the original data archives.

First, the daily weather descriptions are not formalized in the original data archives owing to some input errors in the database. Given the absence of standardized weather condition formats, weather conditions have been described unsystematically, leading to various inconsistencies across various writers and periods. To give an example, the records for the same weather condition can be described using different characters and terms. Specifically, both weather description "或雨或霽" and "或雨或晴" imply the weather condition "rainy and sunny." As the weather category in the original data archives is inordinate, the recorded weather information cannot be directly applied for further investigations.

Second, the original datasets provided the date information only in terms of the traditional lunisolar calendar. For this reason, we converted these reported dates, written in the traditional calendar system of pre-20$^{th}$ century Korea to corresponding dates in the Gregorian calendar, on which modern academic literature rely upon, as a matter of data consistency. Note that our study foremost aimed to establish a database that was readable for scientific purposes. To this end, we identified weather information in the records and compiled this information into pre-defined weather categories.

---

[2] http://sjw.history.go.kr/main.do
[3] https://www.data.go.kr/data/15064218/fileData.do.

**Weather Classification**

To alleviate the aforementioned pitfalls, we first thoroughly defined a series of weather tokens by extracting weather records from the text. Table 1 summarizes the weather tokens for each weather condition. We also included homophonic Chinese characters and interchangeable characters for specific weather conditions. For example, the weather condition 晴 (sunny) could be alternatively expressed as 晴 (sunny) or 清 (clear). All conditions falling into a single category, but expressed by different characters, were thus merged into the same weather token. The defined tokens also contained some indirect weather information in the record.

**Table 1.** Related tokens for weather conditions.

| Weather | Tokens |
|---|---|
| Sunny | 晴, 晴, 清 |
| Cloudy | 陰, 蔭, 雲, 霧, 雷 |
| Rainy | 雨, 兩 |
| Snow | 雪 |
| Hail | 雹 |

Furthermore, we generated unique identifiers for different combinations of weather conditions. More specifically, each weather condition was attributed to different values in terms of the power of 2. For each weather description, we were able to identify a corresponding weather by combining the occurrence of weather conditions. For instance, Figure 2 shows that the weather description "或陰或晴," which indicated the occurrence of "sunny" and "cloudy" weather, based on which the value of weather condition "Sunny and Cloudy" was calculated as 6. Similarly, the weather description "雨雪交下," indicates the occurrence of "rainy" and "snow" conditions, based on which the value of weather condition "Rainy and Snow" was calculated as 24. Through this methodology, each combination of weather conditions was attributed a unique value. Note that such an approach can bundle different descriptions together to the same weather condition. For instance, the expressions "或雨或雹" and "或雨或晴" both reflect rainy and sunny conditions. Thus, such expressions should be categorized as the same weather condition despite different descriptions from the point of view of the characters used. Overall, such an approach can process multiple combinations of weather conditions. More specifically, as each combination should be attributed to a unique value, even multiple combinations of weather conditions such as "朝陰雨晚晴," reflecting that the weather changed from cloudy/rainy to sunny, can be independently categorized.

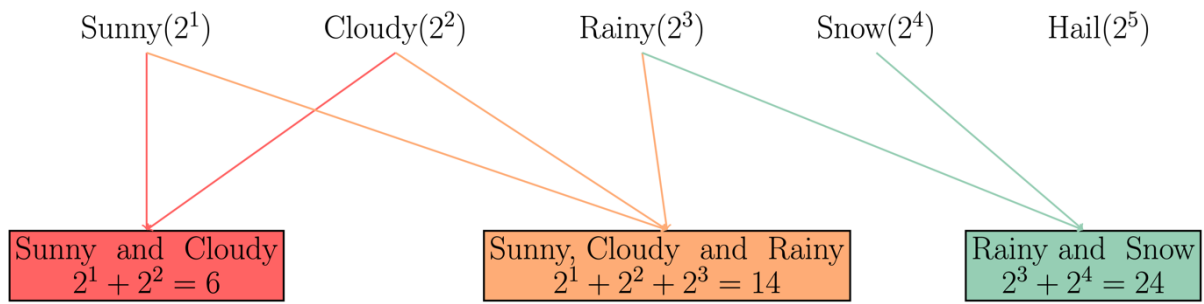

**Figure 2.** Computing Unique Value for Different Combinations of Weather Conditions.

Moreover, we identified a few days with multiple records and weather descriptions, which somewhat hindered the weather record compilation. However, as the proposed method enabled us to extract weather information from relatively complex weather descriptions, we simply merged the multiple weather descriptions and identified the weather information identification and categorization process at the next step.

**Record Dates**

As mentioned, *Seungjeongwon Ilgi* dates each entry in the traditional lunisolar calendar (Lee et al., 2012; NIKH, 1977). Thus, we converted the reported dates to those in the Gregorian calendar. This step was required to convert the analyzed dataset into the contemporary format widely accepted in most research domains including meteorology and climatology.

To this end, we utilized the Python library *sxtwl* [4] for date conversion. In particular, leap years and common years were thoroughly distinguished during date conversion. Moreover, we applied API, provided by the *Korea Astronomy and Space Science Institute* [5] to ensure the validity of the ultimate date conversion. The transformation reliability reference is provided in Table 2, where we summarize several conversion examples. As seen here, the original date in common years and leap years were both accurately converted.

**Table 2.** Example of date conversion: the original date is the lunar date in the original dataset, the converted date is that in the Gregorian calendar.

| Original Date | Converted Date |
|---|---|
| 1623-10-01L0 | 1623-10-24 |
| 1623-10-01L1 | 1623-11-22 |

---
[4] https://github.com/skydancep/sxtwl
[5] https://astro.kasi.re.kr/life/pageView/8

| 1751-05-18L0 | 1751-6-11 |
| 1751-05-18L1 | 1751-7-10 |
| 1870-10-13L0 | 1870-11-5 |
| 1870-10-13L1 | 1870-12-5 |

**Dataset Location and Format**

The daily weather datasets of *Seungjeongwon Ilgi* are publicly available at the Zenodo (doi: 10.5281/zenodo.8142701). This allows any scholar to access these daily weather records. Note that the proper citation of this article and acknowledgment of this database are required for their use, as specified in the end of the current study. Overall, the database provides 100,620 days of weather records with metadata descriptions. The primary information in our data files is organized as follows:

- *ID*: The unique identifier of a specific record in the metadata, which also represents an identifier to merge with external data in the NIKH digital database.
- *Original Date*: The original lunar dates in the NIKH digital database, which are listed in the data format "YYYY-MM-DD". More specifically, "L0" means a leap year and "L1" means a common year.
- *Leap*: The identifier of a leap year.
- *Solar Date*: The Gregorian date converted from the original lunar date. According to the solar calendar, the data cover the following period: April 11, 1623 to August 29, 1910.
- *Weather Text*: The text, describing weather conditions, where multiple weather descriptions of the same day were collocated.
- *Flag*: The computed value, reflecting different combinations of weather conditions. Table 3 summarizes the computed value, the corresponding weather conditions, and the sample size for each weather condition.
- *Volume:* The volume of text in the original record.
- *Herbal Volume:* The volume of text in the herbal record.
- *Sunny:* A dummy variable for "*Sunny*" token. "1" ("0") represents that the weather condition (does not) contains "Sunny" tokens in Table1.
- *Cloudy:* A dummy variable for "*Cloudy*" token. "1" ("0") represents that the weather condition (does not) contains "Cloudy" tokens in Table1.

- *Rainy:* A dummy variable for "*Rainy*" token. "1" ("0") represents that the weather condition (does not) contains "Rainy" tokens in Table1.

- *Snow:* A dummy variable for "*Snow*" token. "1" ("0") represents that the weather condition (does not) contains "Snow" tokens in Table1.

- *Wind:* A dummy variable for "*Wind*" token. "1" ("0") represents that the weather condition (does not) contains "Wind" tokens in Table1.

**Table 3.** Computed value and corresponding weather condition.

| Computed Value | Weather Condition | $N$ |
|---|---|---|
| 2 | Sunny | 76,227 |
| 4 | Cloudy | 8,242 |
| 6 | Sunny and Cloudy | 2,514 |
| 8 | Rainy | 8,753 |
| 10 | Sunny and Rainy | 2,338 |
| 12 | Cloudy and Rainy | 1,322 |
| 14 | Sunny, Cloudy, and Rainy | 100 |
| 16 | Snow | 701 |
| 18 | Sunny and Snow | 366 |
| 20 | Cloudy and Snow | 244 |
| 22 | Sunny, Cloudy, and Snow | 18 |
| 24 | Rainy and Snow | 134 |
| 26 | Sunny, Rainy, and Snow | 7 |
| 28 | Cloudy, Rainy, and Snow | 11 |
| 32 | Hail | 2 |
| 40 | Rainy and Hail | 10 |
| 42 | Sunny, Rainy, and Hail | 6 |
| 44 | Cloudy, Rainy, and Hail | 1 |
| 46 | Sunny, Cloudy, Rainy, and Hail | 2 |
| 56 | Rainy, Snow, and Hail | 1 |
| 62 | Sunny, Cloudy, Rainy, Snow, and Hail | 1 |

We utilize modern weather data[6] to further validate our weather condition detection method. Specifically, we selected two cities, Incheon and Seoul, which are adjacent to the observation points of the *Seungjeongwon Ilgi*, and utilized overlapping meteorological data to conduct the validation analysis. Figure 3-A compares the rainfall amounts in Incheon from 1904 to 1908 across different weather condition categories detected from the *Seungjeongwon Ilgi* on the according date. We find that compared to cloudy, rainy, and snowy days, the days recorded as sunny in the *Seungjeongwon Ilgi* had significantly lower amounts of rainfall. As depicted in

---
[6] The modern weather data is obtained from Open MET Data Portal. The available data includes rainfall amounts in Incheon from 1904 to 1908, rainfall amounts and duration of sunshine in Seoul from 1908 to 1910.

Figure 3-B, the similar pattern can also be found in the comparison of the rainfall amounts in Seoul from 1908 to 1910 across different weather condition categories. Additionally, Figure 3-C compares the duration of sunshine in Seoul from 1908 to 1910 among different weather condition categories. Clearly, compared to cloudy, rainy, and snowy days, the days recorded as sunny in the *Seungjeongwon Ilgi* have a longer duration of sunshine with the average value of 7.14 hrs. Taken together, although *Seungjeongwon Ilgi* cannot provide detailed information about weather conditions like modern weather observation, these validations serve to demonstrate that our weather detection method can approximate the weather conditions of the corresponding dates.

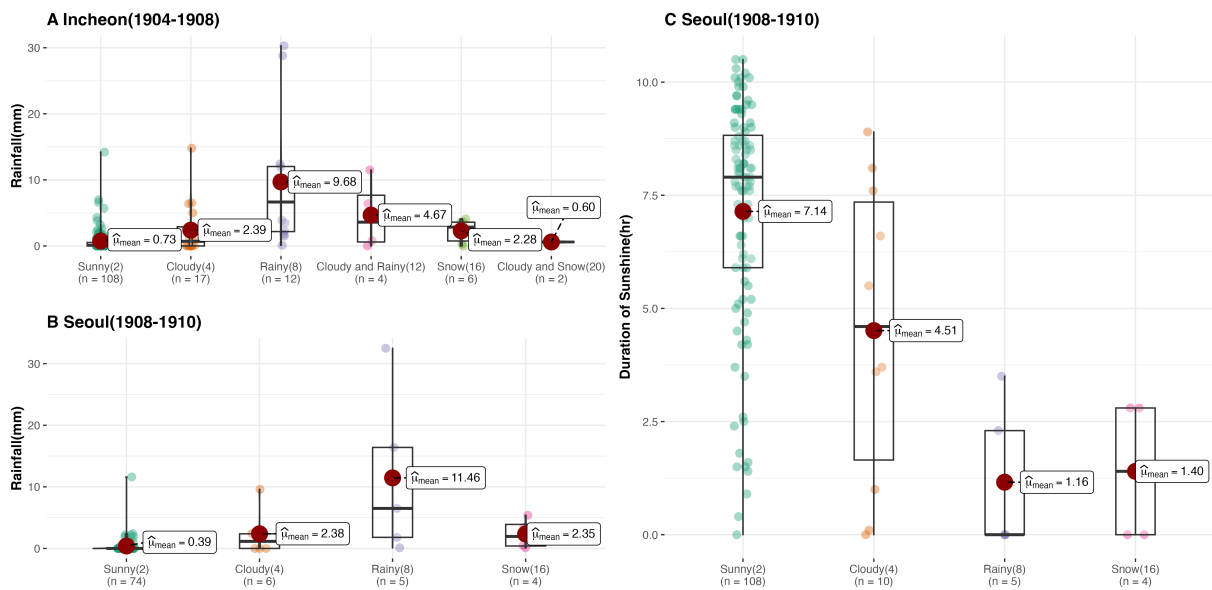

**Figure 3.** Validation of weather condition detection using modern weather condition data. (A) Comparison of rainfall amounts in Incheon across different weather categories detected from the *Seungjeongwon Ilgi*. (B) Comparison of rainfall amounts in Seoul across different weather categories detected from the *Seungjeongwon Ilgi*. (C) Comparison of duration of sunshine in Seoul across different weather categories detected from the *Seungjeongwon Ilgi*.

**Data Gaps and Lost Data**

Although the NIKH dataset and our proposed data processing method are very promising, some limitations should be discussed owing to the data gaps in our dataset. First, some parts of the records have been destroyed or lost. Second, some other parts of the records did not refer to any weather conditions, thereby creating inevitable gaps in the reporting of weather on the corresponding dates. Third, several weather descriptions could not be identified into the pre-defined weather category weather description. It should be noted that a part of the records contains the description of "wind", which is of great importance in meteorological analysis. However, we find that description of "wind" is somehow a particular style of record by the specific observers. Since the pre-defined weather category is intended to include the combinations of weather descriptions that are broadly recorded by different observers across times, the description of "wind" is not included in the pre-defined weather category. Instead, for researchers who are interested in the description of "wind", we provide an additional dummy variable for identifying these descriptions.

Figure 4 shows the missing weather categories in years when the number of missing days exceeded 30. Specifically, the red block reflects the missing weather categories owing to the absence of records, while the blue block indicates the missing weather categories due to the absence of weather information in the records or the presence of unidentified weather information. Note that a remarkably large number of records were absent in 1624 and 1695.

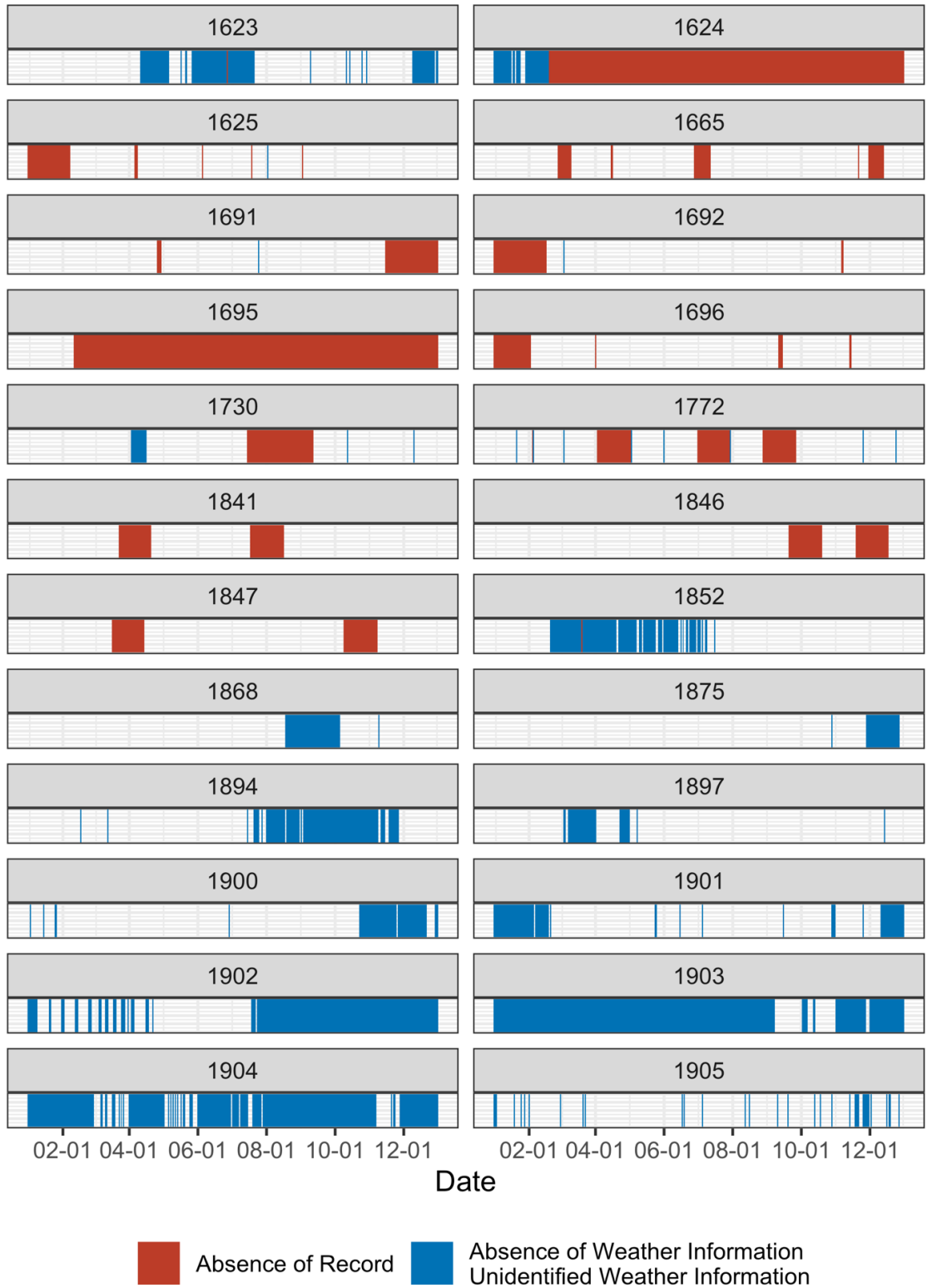

**Figure 4.** Missing Record and Missing Weather Information in the NIKH Dataset.

These gaps can be explained in many cases as *Seungjeongwon Ilgi* has been partially burned and renovated several times in its history. In principle, *Seungjeongwon Ilgi* was recorded in chronological order, with one volume per month, but two or more volumes were produced at times of such incidents. Parts of the *Seungjeongwon Ilgi*, which were written before

*Gwanghaegun*, the 15th ruler of the *Joseon* Dynasty in the early 17th century, were destroyed by fire and invasion. As a result, only the records, written after King *Injo* remained until modern days. The original diaries spanned 3,243 volumes and covered 288 years in the latter half of the *Joseon* Dynasty from the first year of King *Injo* (1623) to the fourth year of *King Soonjong* (1910). These original diaries are now preserved in the *Kyujanggak* Institute for Korean Studies of Seoul National University. Of these, > 900 volumes were supplemented after being burned. Notably, *Seungjeongwon Ilgi* was stored at *Seungjeongwon* in the *Gyeongbokgung* Palace for ~200 years during the early *Joseon* Dynasty. However, the records subsequently burned down alongside other national treasures during the *Imjin* War in the 25th year of King *Seonjo* (1592). Moreover, *Seungjeongwon Ilgi* for the next 32 years (until the first year of King *Injo* in 1623) also burned down during the *Yi Gwal's* Rebellion in the 2nd year of King *Injo* (1624). Following these events, the records were supplemented with 26 volumes of *Seungjeongwon Ilgi*. As a result, an official gazette, well preserved by *Hong Deok-lin*, a *Seungjeongwon* scribe, has been kept since then in the east side of the *Injeongjeon* of the *Changdeokgung* Palace. However, in the 20th year of King *Yeongjo* (1744), a fire broke out in *Seungjeongwon* on the 13th day of the 10th month, once again destroying considerable records in *Seungjeongwon Ilgi* (*Yingjo sillok*). Because of this, in the 5th month of the 22nd year of King *Yeongjo* (1746), a diary office *Ilgicheong* was established. According to *Yingjo sillok* and the transcript of the diary office, 1,796 volumes of *Seungjeongwon Ilgi* covering 130 years were burned down. Note that by the end of 1747, 548 volumes were restored, but they constituted only one-third of the original collection, which had been previously burnt (Nakamura, 1965). The continuous lack of records for the years 1624, 1625, 1665, 1691, 1692, 1695, 1696, and 1730 in Figure 4 is assumed to be due to this fire. This absence includes not only weather data but also all other records, such as the king's commands, papers, and occurrences. In the 25th year of King *Gojong* (1888), a fire broke out in *Seungjeongwon* again and 361 volumes from the 2nd year of King *Cheoljong* (1851) to the 25th year of King *Gojong* (1888) were destroyed (*Gojong sillok*). This fire caused the weather data in Figure 4 for the years 1852, 1868, 1875, 1894, and 1897, to be missing. Besides these fires, *Seungjeongwon Ilgi* has been lost or destroyed several times, and repaired subsequently. Moreover, the vast data loss in the 1900s coincided with the Japanese invasion in Korea, exacerbated by the instability of the political situation in Korea. These parts have never been restored. It should be emphasized that the absence of weather reporting just indicates that the weather information was not recorded, not a continual absence of a certain weather. Since there was no significant drought or floods during the year with missing data, any subsequent study will not be impacted.

**Summary and Discussion**

This study utilized daily weather records of *Seungjeongwon Ilgi* from the NIKH database. *Seungjeongwon Ilgi* is a daily record of the *Seungjeongwon*, the Royal Secretariat of the *Joseon* Dynasty of Korea. These diaries cover the period of 1623–1910, with many records, including data on the daily weather. The records reported weather conditions of the location at N 37°35′, E 126°59′, corresponding to the modern Seoul urban area. We used the weather records from the NIKH database and classified the daily weather using a text mining method. We also converted the report dates from the traditional lunisolar calendar to the Gregorian calendar, to contextualize the data into the contemporary daily measurements and to the widely accepted format for meteorological measurements. Our resultant datasets are publicly available at doi: 10.5281/zenodo.8142701; the citation of our study and acknowledgment to the data source are both required.

We have also visualized the data and identified several data gaps, which were likely caused by the numerous fires and political unrest episodes. To this end, Figure 5 illustrates the proportion of missing records, missing weather descriptions, and unidentified weather information. As seen, the missing data cluster accounts for a somewhat minor segment of the entire span of weather records from *Seungjeongwon Ilgi*. Thus, we concluded that our dataset includes continuous and systematic weather records that can provide valuable insights into the long-term historical climate changes in the Korean Peninsula. Our datasets include 288 years chronologically, thereby uniquely covering a major part of the 17$^{th}$ century crisis. During this crisis, numerous social unrest events, plague pandemics, climate anomalies, and solar grand minima occurred, as previously reported (Hayakawa, Lockwood, et al., 2021; Parker, 2013; Soon and Yaskell, 2013). Overall, our proposed data offer a unique reference for the daily weather in the Korean Peninsula back to the early 17$^{th}$ century, extending to 1910.

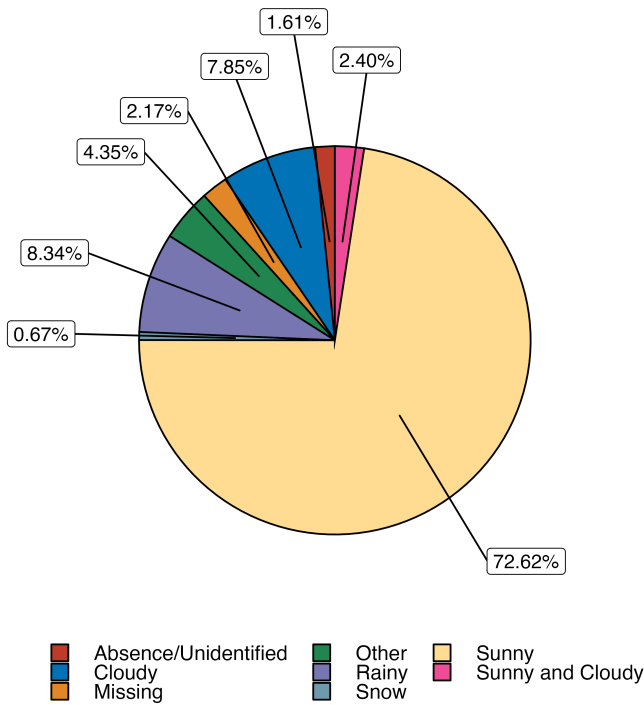

**Figure 5.** Distribution of Weather Categories: we only list the main category while other weather categories are included in "Other." Especially, "Absence/Unidentified" indicates missing weather information owing to the absence of weather information in the records or the presence of unidentified weather information. "Missing" indicates missing weather information owing to the absence of a record.

**Acknowledgements**

This work is supported by Program for Establishing a Consortium for the Development of Human Resources in Science and Technology, Japan Science and Technology Agency (JST) and is supported by FRIS Creative Interdisciplinary Collaboration Program (K. Ichikawa). KI also acknowledges Young Researchers Ensemble Grants 2022 (Tohoku University) and Ensemble Continuation Grants for Early Career Researchers 2022 (Tohoku University). HH thanks young researcher units for the advancement of new and undeveloped fields, Institute for Advanced Research, Nagoya University of the Program for Promoting the Enhancement of Research Universities, JSPS Grant-in-Aid 21K13957, JSPS Overseas Challenge Program for Young Researchers, the ISEE director's leadership fund for FY2021, and Young Leader Cultivation (YLC) program of Nagoya University. HH and YK thank the young researcher units for the advancement of new and undeveloped fields, Institute for Advanced Research, Nagoya University of the Program for Promoting the Enhancement of Research Universities.